\documentclass{ws-procs9x6}

\newcommand{\nn}{\nonumber}
\newcommand{\be}{\begin{equation}}
\newcommand{\ee}{\end{equation}}
\newcommand{\ba}{\begin{eqnarray}}
\newcommand{\ea}{\end{eqnarray}}
\newcommand{\ci}[1]{\cite{#1}}
\def\vk{{\bf k}_{\perp}}

\def\vb0{{\bf b}_0}

\def\als{\alpha_s}

\def\gev{\,{\rm GeV}}

\def\xbj{x_{\rm Bj}}

\newcommand{\wf}{wavefunction}

\newcommand{\gsim}{\raisebox{-4pt}{$\,\stackrel{\textstyle
                                                         >}{\sim}\,$}}

\newcommand{\tw}{\textwidth}
                                                         
\newcommand{\req}[1]{(\ref{#1})}
\def\xb{\bar{x}}

\def\={\,=\,}

\begin{document}

\title{HARD EXCLUSIVE SCATTERING AT JLAB}

\author{P. KROLL}

\address{Fachbereich Physik, Universitaet Wuppertal,\\
D-42097 Wuppertal, Germany\\
$^*$E-mail: kroll@physik.uni-wuppertal.de}

\begin{abstract}
The various factorization schemes for hard exclusive processes and
the status of their applications is briefly reviewed. 

\noindent Invited talk presented at the workshop on Exclusive Processes at High 
Momentum Transfer, Newport News, Virginia USA, May 2007
\end{abstract}


\bodymatter
\section{Introduction}\label{sec1}
In large momentum transfer exclusive processes the probe, say, a
virtual photon has a wave length that is much shorter than the spacial 
extension of the hadronic target. This allows to look inside the
hadrons and to study the interactions of their constituents, quarks and
gluons. There is overwhelming evidence, mainly from inclusive
reactions, that QCD is the correct theory for the interactions between
quarks and gluons. QCD is a complicated theory. Quarks and gluons are 
confined, only their bound states - the hadrons - can be observed 
experimentally. The formation of hadrons from quarks and gluons occurs 
at soft scales where QCD perturbation theory is inapplicable. But,
with the exception of lattice QCD, there is no analytical or numerical 
method known to solve QCD in the soft region. In any scattering process 
as hard as the external scale, for instance the virtuality, $Q^2$, of 
the probing photon, may be, soft hadronization is unavoidably involved
too. Thus, one may wonder whether it is possible to calculate
observables for hard processes. This is indeed possible in a number of 
cases thanks to the factorization properties of QCD:  hard processes 
factorize into parton-level subprocesses amenable to perturbative QCD 
(and/or QED), and in soft hadronic matrix elements which embody the 
non-perturbative physics. For a number of processes there are rigorous
proofs of factorization available, e.g.\ the pion electromagnetic form 
factor, deeply virtual lepton-nucleon scattering (DIS), deeply virtual 
Compton scattering (DVCS). For others factorization is a hypothesis
with often good  arguments for its validity. However, we have to be
cautious in these cases. Collins and Qiu \ci{collins07} found a 
counterexample, namely $h_1 h_2\to h_3 h_4 X$ where $h_i$ denotes a 
hadron, for which ($\vk$) factorization breaks down. Given the 
theoretical complications involved in exclusive scattering and with 
regards to the large number of succesful tests of QCD properties 
accumulated over the last 30 years, the experimental and theoretical 
investigation of hard exclusive processes will not contribute towards 
the verification of QCD, rather we will learn about methods how to
apply QCD.
 
In the following I will briefly review the factorization schemes used 
in exclusive scattering (Sects.\ \ref{sec2} and \ref{sec3}). In Sect.\ 
\ref{sec4} I will summarize our present knowledge on the
generalized parton distributions (GPDs), the soft hadronic matrix 
elements occuring in the handbag factorization scheme. 
Next I will turn to applications of the handbag factorization to 
deeply virtual exclusive scattering (Sect.\ \ref{sec6}), discuss 
alternative theoretical approaches such as the Regge model (Sect.\ 
\ref{sec7}), and turn finally to wide-angle exclusive reaction (Sect.\ 
\ref{sec8}). Special emphasis is laid on the role of JLab in this 
physics - what has been achieved by JLab till now, what will be done 
in the future. In Sect.\ \ref{sec9} I will present the summary.

\section{The ERBL factorization scheme}\label{sec2}
A first factorization scheme for hard exclusive processes has been
invented around 1980. Efremov and Radyushkin \ci{efremov89} as well as 
Brodsky and Lepage \ci{bro89} showed that factorization holds for the
pion form factor at large $Q^2$. This factorization scheme has been 
generalized later on to many other exclusive processes, lacking
however proof in most cases~\footnote{Many authors have also
  contributed to the development of that field, e.g.\ Refs.\
  \refcite{duncan,chernyak}.}.
Since the evolution equation for the associated soft matrix element, 
the so-called distribution amplitude (DA), is named after these
authors, I take the liberty to give the full factorization scheme also
this name - ERBL factorization. Other frequently used names for it are 
either misleading or lead to a clash of notation.  
 
\begin{figure}
\begin{center}
\psfig{file=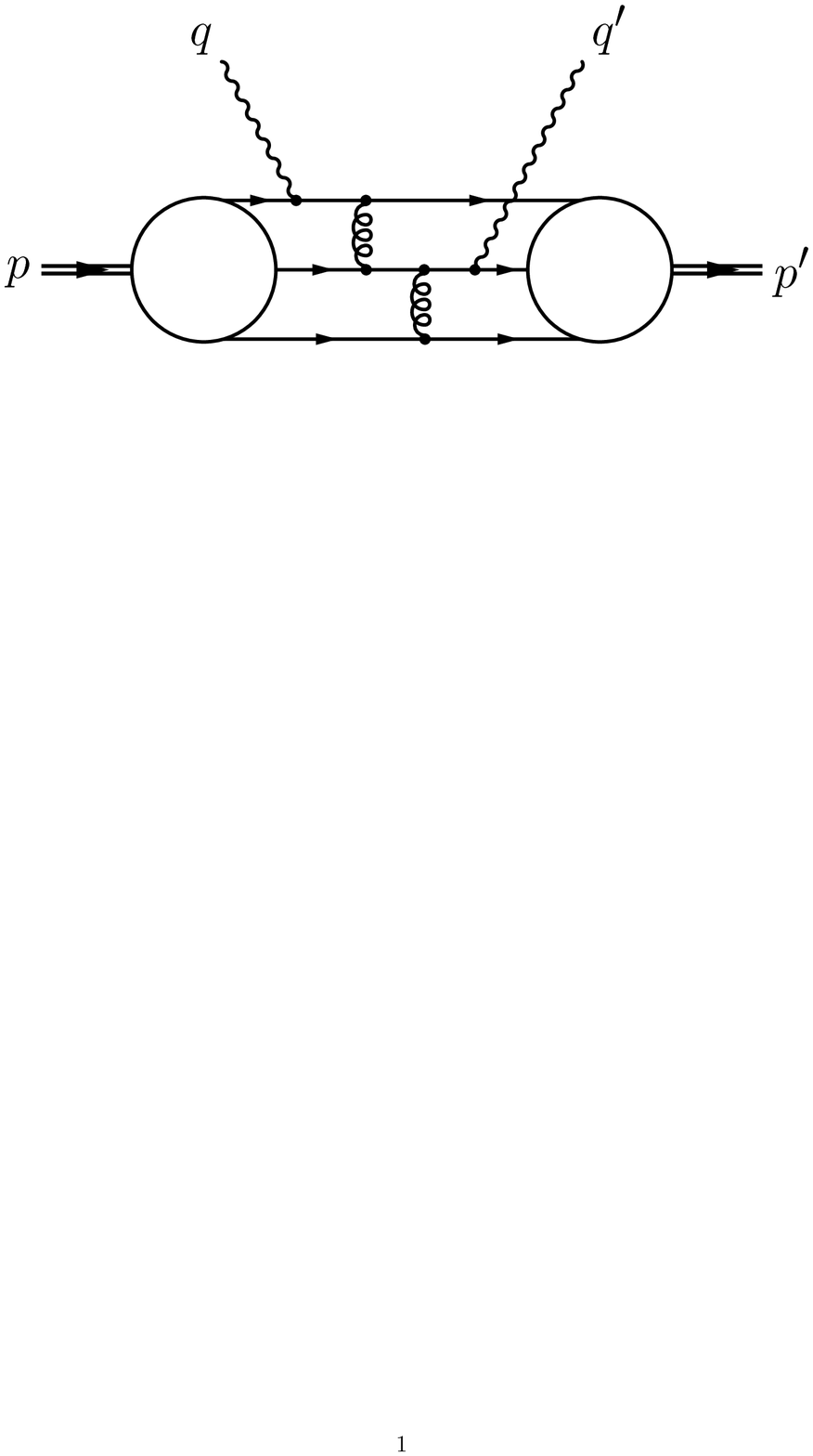,bb=90 532 510 795,width=0.44\tw}\hspace*{2em}
\psfig{file=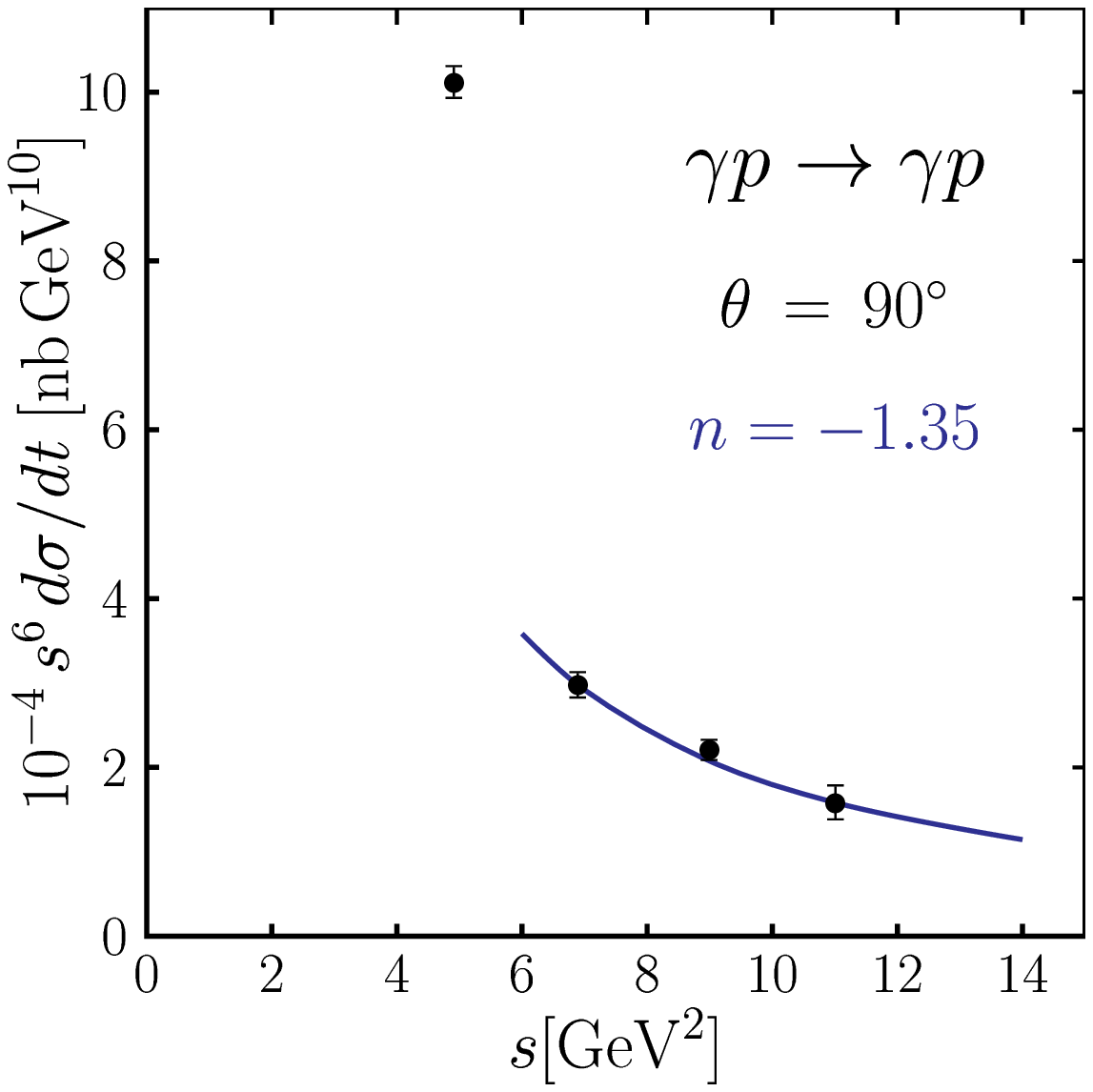,bb=114 390 448 725,width=0.32\tw}
\end{center}
\caption{Left: A typical graph for Compton scattering within the ERBL
  factorization scheme. Right: The Compton cross section, scaled by
  $s^6$, at a scattering angle of $90^\circ$. Data taken from Ref.\
  \refcite{halla-compton}.}
\label{fig:compton-graph}
\end{figure}

In order to sketch the ERBL scheme let me consider Compton scattering
off protons at large Mandelstam variables $s, -t, -u$ as a typical and
important example and let me consider only the at large scales dominant 
valence Fock state of the proton. The amplitudes of this process 
factorize into partonic subprocess $\gamma qqq\to \gamma qqq$ (see 
Fig.\ \ref{fig:compton-graph}) and in proton DAs $\Phi_p(x_1,x_2,x_3)$ 
where the $x_i$ are the usual momentum fractions. All partons of the 
valence Fock state participate in the subprocess, they are
emitted or absorbed collinearly from their parent hadron and are quasi 
on-shell. This neccessitates the exchange of at least two hard gluons. 
The Compton amplitudes are given by convolutions of subprocess 
amplitudes and DAs
\be
M \sim \Phi_p \otimes {\cal H} \otimes \Phi_p\,.
\ee
One may also consider higher Fock states of the involved hadrons but
these contributions are suppressed by inverse powers of the hard scale
as compared to the valence Fock state contribution. 

The ERBL factorization scheme implies dimensional counting
\ci{matveev,farrar} which means that at large momentum scales (or
short distances) exclusive observables exhibit scaling, i.e.\ the fall
off as a certain power of the hard scale asymptotically. The power 
laws are modified by perturbative logs generated by the running of 
$\als$ and the evolution of the DAs. Scaling often holds approximately 
in experiment although there seems to be no evidence for the
perturbative logs. Recent precision data are often in conflict with 
dimensional counting. As an example the JLab Hall A data
\ci{halla-compton} on Compton scattering are shown in Fig.\
\ref{fig:compton-graph}. Clearly the cross section does not drop as 
$s^{-6}$ as predicted by dimensional counting. Violations of
dimensional counting are also seen in the Pauli form factor \ci{pauli} 
or in the precise BELLE data \ci{belle} on $\gamma\gamma\to p\bar{p}$. 
These counterexamples do not disprove dimensional counting. They
merely indicate that the experimentally available scales for these
data are not sufficiently large for applying dimensional counting and, 
hence, the ERBL factorization scheme.

The ERBL factorization scheme has been frequently applied to various
exclusive processes, e.g.\  electromagnetic form factors, Compton 
scattering, photoproduction of mesons and various time-like
processes. It turned out however that with very few exceptions the 
size of the ERBL contribution is too small, often by order of
magnitude, in comparison with experiment. What does this mean?  
Are the scales available in present-day experiments, typically
about $10\,\gev^2$, too low for applying ERBL factorization or is it
possible to improve the results within that scheme? For instance, one
may follow the suggestion by Chernyak and Zhitnitky \ci{chernyak}
and use DAs which are concentrated in the end-point regions where one
of the momentum fractions tends to zero. Such DAs provide much larger ERBL 
contributions, in some cases even agreement with experiment is
achieved, e.g.\ for the pion form factor. It has been argued that the
use of these CZ-type DAs lead to theoretical inconsistencies
\ci{isgur84,radyuskin84} since the bulk of the perturbative
contribution is accumulated in the end-point regions where
perturbation theory breaks down. One may also suspect that higher
order pQCD corrections lead to a large K-factor but this has not yet
been elaborated. However, the known NLO corrections for the pion form 
factor do not suffice for solving the difficulties with the size of
the ERBL contribution if evaluated from DAs close to the asymptotic
one, $\Phi_{AS}=6x(1-x)$.

In order to cure some of the deficiencies of the ERBL
factorization scheme Sterman and Li \ci{sterman93} invented to
so-called modified perturbative approach in which the quark transverse
momenta are retained and Sudakov suppressions are taken into account.
Configurations with large transverse separations of the quarks which
occur in the end-point regions are suppressed and theoretically
consistent results are obtained. For consistency the DAs are to be
replaced by transverse momentum dependent light-cone \wf
s \ci{jakob93}. In general the contributions obtained with
the modified perturbative approach are also too small even if CZ-type \wf
s are used, see for instance Ref.\ \refcite{stefanis95}.
 
\section{Handbag factorization in exclusive reactions}\label{sec3}
\begin{figure}
\begin{center}
\psfig{file=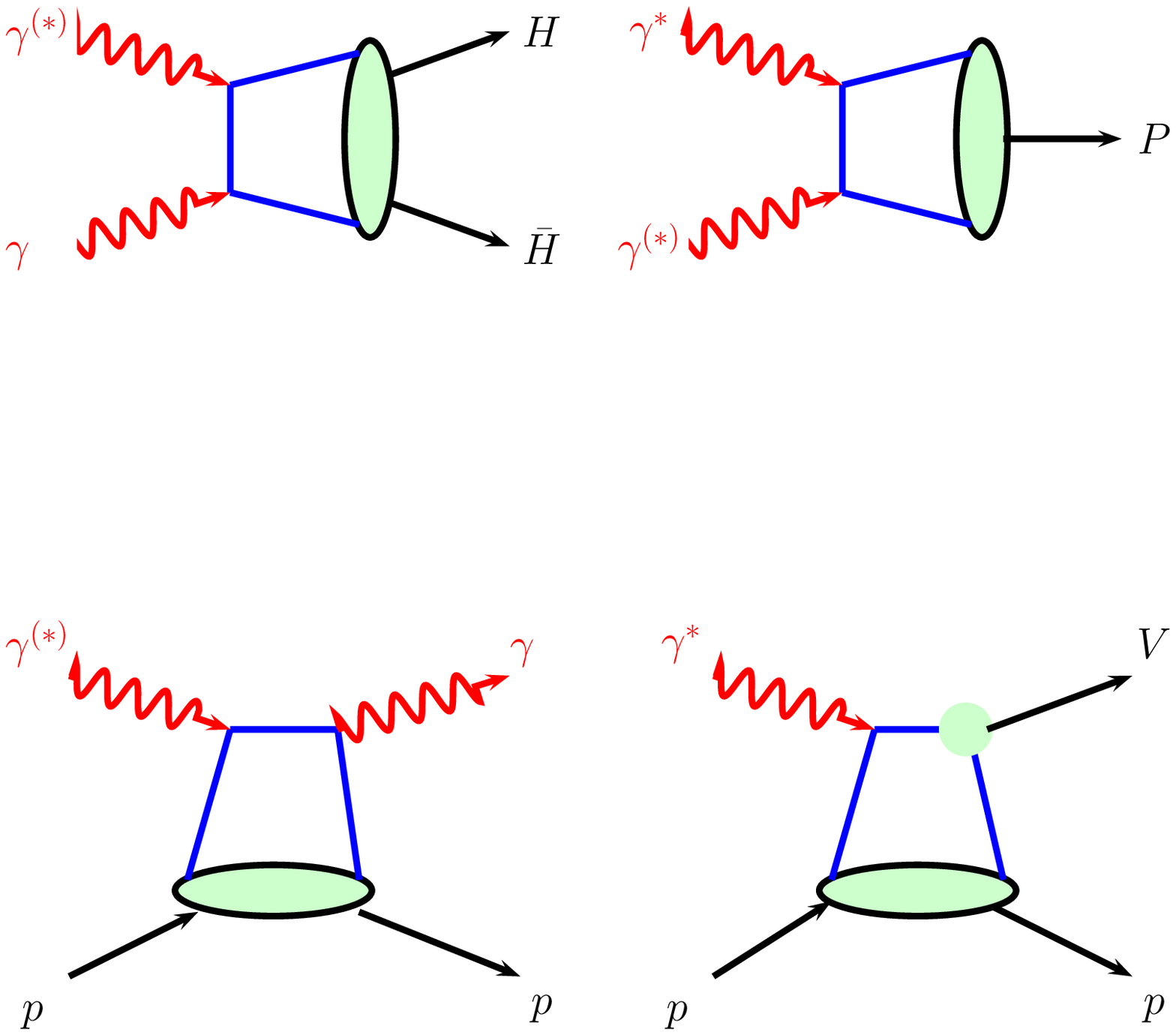,bb= 120 238 345 407,width=0.30\tw,clip=true}
\psfig{file=handbag-c.ps,bb=115 505 590 647,width=0.65\tw,clip=true}
\end{center}
\caption{Handbag factorization in the space- and time-like regions
  and the form factor for photon-pseudoscalar-meson transitions.}
\label{fig:handbag-graph}
\end{figure}

A new factorization scheme \ci{mueller94,rad96,ji96} became popular in
1996. In constrast to the ERBL scheme there is only one active parton
that participates in the partonic subprocess, e.g.\ for Compton
scattering $\gamma q \to \gamma q$, see Fig.\ \ref{fig:handbag-graph}. 
Similar to the ERBL scheme the active parton is emitted and reabsorbed
by the hadron collinearly and is quasi on-shell. The soft hadronic
matrix elements are now GPDs. The handbag factorization applies to two 
different kinematical regions of exclusive reactions. The deeply
virtual region is characterized by large $Q^2$ but small Mandelstam $-t$.
In the wide-angle region, on the other hand, $Q^2$ is assumed to be
small while $-t$ and $-u$ are considered as large. Reactions studied
in both the regions are Compton scattering and photo- and
electroproduction of mesons in the space-like region as well as the 
crossed processes (e.g.\ $\gamma^{(*)}\gamma\leftrightarrow p\bar{p}$, 
$p\bar{p}\to \gamma^{(*)} M$) in the time-like region (see Fig.\ 
\ref{fig:handbag-graph}). Related to these processes is the 
photon-pseudoscalar-meson (P) transition form factor (see Fig.\ 
\ref{fig:handbag-graph}). The partonic subprocess is identical to that
of Compton scattering but the hadronic matrix element for the
$q\bar{q}\to P$ transition is just the DA appearing in the ERBL
factorization scheme. The transition form factor is an exceptional
case since the handbag and the ERBL factorization schemes fall
together for it. The theoretical result for it, say, to NLO and 
evaluated from the asymptotic $\pi$ DA is very close to experiment
\ci{savinov}.  Only the origin of the remaining about $10\%$ is still 
under debate. Suggested have been NNLO corrections, deviations from 
the asymptotic DA and/or power corrections.  
  
For parton helicity non-flip there are four GPDs for the proton in the
space-like region denoted by $H$, $\widetilde{H}$, $E$ and
$\widetilde{E}$. They exist for each quark flavor and for the gluon and 
are functions of three variables, a momentum fraction $x$, the skewness 
$\xi$ and $t$. For the GPDs a number of properties are known. Thus, $H$ 
and $\widetilde{H}$ reduce to the ordinary unpolarized  and polarized
parton distribution functions (PDFs) in the forward limit $\xi, t\to 0$
\ba
H^q (x,0,0) &=&\, q(x)\,, \qquad \;\, \widetilde{H}^q(x,0,0) =\, \Delta
q(x)\,,\nn\\
H^g (x,0,0) &=& xg(x)\,, \qquad \widetilde{H}^g(x,0,0) = x\Delta
g(x)\,.
\label{reduction}
\ea
The forward limits of $E$ and $\widetilde{E}$ are not accessible in
DIS. The GPDs are related to the proton form factors by sum rules, e.g.\ 
for the Dirac form factor
\be
F_1^a(t)\=\int_{-1}^1 dx H^a(x,\xi,t)\,, \qquad
F_1(t)\=\sum_a\,F_1^a(t)\,.
\label{sum-rules}
\ee
Analogous sum rules for $E$ being related to the Pauli form factor,
$\widetilde{H}$ (related to the axial form factor) and $\widetilde{E}$
(related to the pseudoscalar form factor) hold. Other known properties of
the GPDs are polynomiality, universality, evolution, Ji's sum rule
and a couple of positivity constraints. 

One may also consider parton helicity flip. These configurations
define four more GPDs, termed $H_T$, $\widetilde{H}_T$, $E_T$ and 
$\widetilde{E}_T$ for each quark flavor and for the gluon. These 
functions are practically unknown. They are very hard to access 
since parton helicity flip is frequently suppressed in partonic 
subprocesses. One may proceed and consider two (or more) active
partons. It is straightforward to show that in order to match the 
requirement of collinear emission and absorption of quasi on-shell 
partons by the hadrons, at least one hard gluon is to be exchanged 
between the active partons. These contributions which have not yet 
been investigated, are therefore expected to be suppressed. It is 
interesting to note that, say, for Compton scattering off protons the 
case of three active partons is just the ERBL contribution if 
dominance of the valence Fock state is assumed. 

\section{What do we know about the GPDs?} \label{sec4}
A popular model which allows to construct the GPDs from the PDFs is 
the double distribution ansatz \ci{radyushkin} 
\be
f_i(\beta,\alpha,t)\= g_i(\beta,t)\,h_i(\beta)\,
                   \frac{\Gamma(2n_i+2)}{2^{2n_i+1}\,\Gamma^2(n_i+1)}
                   \,\frac{[(1-|\beta|)^2-\alpha^2]^{n_i}}
                           {(1-|\beta|)^{2n_i+1}}\,.
\label{DD}
\ee 
The functions $h_i$ represent the PDFs. In the case of $H$ for instance 
\be
h_g\= |\beta|g(|\beta|)\,,\;
h^q_{\rm sea} \= q_{\rm sea}(|\beta|)\;{\rm sign}(\beta)\,,
\;                      
h^q_{\rm val} \= q_{\rm val}(\beta)\, \Theta(\beta)\,,    
\label{function-h}
\ee
and $g_i(\beta,t=0)=1$, $n_i$ either 1 or 2. 
The GPD is obtained from the double distribution by the following
integral representation
\be
H_i(\xb,\xi,t)\=\int_{-1}^1 d\beta\,\int_{-1+|\beta|}^{1-|\beta|} d\alpha\,
                 \delta(\beta+\xi\alpha-\xb)\,f_i(\beta,\alpha,t)\, +
		       {\rm D-term} \,.
\label{GPD-DD}
\ee
For the $t$ dependence of the GPDs, embodied in the function $g_i$,
several ansaetze are to be found in the literature. The simplest idea
is to assume that it represents a $\beta$ independent kind of form
factor but the implied $\beta - t$ factorization seems to be 
unrealistic \ci{DFJK1,goeke}. Another idea is to generalize the Regge
behaviour of the PDFs \ci{landshoff}, $q(\beta)\to \beta^{-\alpha(0)}$ 
for $\beta\to 0$, to non-zero values of $t$:
\be 
g_i(\beta,t)=  {\rm e}^{b_it}\,\mid\beta\mid^{-\alpha_i'\,t}\,,
\ee
Here a linear Regge trajectory, $\alpha_i=\alpha_i(0) +
\alpha'_i\,t$, is assumed and an exponential $t$ dependence of the
corresponding residue (with a parameter $b_i$). There are many
applications of the double distribution model, reggeized or not, for 
instance Refs.\ \refcite{VGG,freund, mueller06,guzey,gk2,diehl05}. 
The advantage of the double distribution model is that the reduction 
formulas \req{reduction} and polynomiality are automatically
satisfied. The $D$-term \ci{weis} which is not related to the PDFs and 
hence a free function, provides the largest power of $\xi$ in the
moments. It only contributes to the real parts of the gluon and the 
flavor-singlet quark GPDs. Its quantitative role is not clear. 

Alternatively, one may  try to extract the GPDs from experimental data
in analogy to the determination of the PDFs. First attempts to
determine at least the zero-skewness GPDs this way have been published
\ci{guidal,DFJK4}. The idea is to exploit the sum rules \req{sum-rules} 
at zero skewness, e.g.\ 
\be
F_{1}^{u} \= \int_0^1 dx\, H^u_v
= 2 F_1^p + F_1^n\,,\quad
F_{1}^{d} \= \int_0^1 dx\, H^d_v
 = 2 F_1^n + F_1^p\,,
\label{eq:sumrules}
\ee
where the valence quark GPDs are defined by $H^{q}_v=H^q - H^{\bar{q}}$.
A possible contribution from $H^s-H^{\bar{s}}$ has been neglected in
\req{eq:sumrules}. The measurements of the strangeness form factors
\ci{g0,HAPPEX,A4} seem to indicate that this contribution
is small although non-zero \ci{thomas}. A weak evidence for
$s(x)\neq \bar{s}(x)$ has been found by the  CTEQ group \ci{cteq07}. 

To determine the integrand from the integral is an ill-posed problem
in a strict mathematical sense. But using an ansatz for the GPDs with a
few free parameters adjusted to experiment, it is possible to extract 
the GPDs $H$, $\widetilde{H}$ and $E$ for valence quarks. Admittedly
the results on the GPDs depend on the ansatz which one may take as
\be
H^q_v= q_v(x) \exp{[f_q(x)t]}\,,
\label{eq:ansatz}
\ee
in which 
\be
f_q = [\alpha'\log(1/x)+ B_q]\,(1-x)^{n+1} + A_qx(1-x)^n\,,
\label{eq:profile}
\ee
and analogously for the other two GPDs. In Ref.\ \refcite{DFJK4} a
standard slope for the Regge trajectory is assumed ($\alpha'=0.9\,\gev^2$), 
$n=2$ taken and the CTEQ6 PDFs \ci{cteq6} are used as input. The parameters
$A_q$ and $B_q$ are fitted to the form factor data. In Ref.\
\refcite{guidal}, on the other hand, $A_q=B_q=0$ is assumed as well as
$n=0$ while $\alpha'$ is fitted to the data. The ansatz
\req{eq:ansatz}, \req{eq:profile} is motivated by overlap of Gaussian 
light-cone \wf s at large $-t$ and large $x$ and by Regge behavior at
low $-t$ and small $x$ (cf.\ the double distribution ansatz \req{DD}
and \req{GPD-DD} in the limit $\xi\to 0$). It should be noted that
there is a third somewhat different attempt to extract the zero-skewness 
GPDs from the form factors \ci{ahmed}.  

The GPDs $H$ and $\widetilde{H}$ extracted from the form factor data
look similar to the corresponding PDFs at low $-t$ while, at larger
$-t$ (beyond the zero of the Regge trajectory), all GPDs exhibit a
pronounced peak which moves towards $x=1$ with increasing $-t$. 
The GPDs $H^u_v$ and $H^d_v$ are both positive while 
$\widetilde{H}^u_v$ and $\widetilde{H}^d_v$ as well as $E^u_v$ and
$E^d_v$ have opposite signs. The double distribution model \req{DD},
\req{GPD-DD} possess also this property. The signs and sizes of the
valence quark GPDs are fixed by the known lowest moments of the GPDs
at $\xi=t=0$ ($e_v^a(x)=E^a_v(x,\xi=0,t=0)$)
\ba
\int_0^1 dx\,u(x) &=& 2\,, \;
\int_0^1 dx\,\Delta u(x) \= \phantom{-}0.926\,, \; 
\int_0^1 dx\,e^u_v(x) \= \phantom{-}1.67\,, \nn\\  
\int_0^1 dx\,d(x) &=& 1\,, \; 
\int_0^1 dx\,\Delta d(x) \= -0.341\,, \;
\int_0^1 dx\,e^d_v(x) \= -2.03\,.
\ea  
The moments of $\widetilde{H}$ are known from $\beta$ decays, those of
$E$ from the anomalous magnetic moments of proton ad neutron. Given
that the GPDs are smooth functions without zeros they should reflect
the properties of the moments at least at low $\xi$ and low $-t$.

With the valence quark GPDs at hand one may evaluate Ji's sum rule 
\ci{ji96} and determine the total and orbital angular momentum the 
valence quarks carry. Thus, for instance, from the GPDs derived in 
Ref.\ \refcite{DFJK4} one obtains
\ba
L^u_v &=& -(0.24\div 0.27)\,,  \qquad J^u_v = \phantom{-}0.21\div 0.24\,, \nn\\
L^d_v &=& \phantom{-(}0.15\div 0.19\,,  \hspace*{0.075\tw} J^d_v = -0.02\div 0.02\,.
\label{angular-momentum}
\ea
The opposite signs of $L^u_v$ and  $L^d_v$ but nearly the same
magnitude are related to the corresponding property of $E_v$.
Fourier transforming the zero-skewness GPDs with respect to the
momentum transfer ${\bf \Delta}$ (${\bf \Delta}^2=-t$) \ci{burkhardt} 
one learns about the transverse localization of partons, i.e.\ about 
their densities in the hybrid representation of longitudinal momentum
fraction and transverse configuration space. One may also evaluate
various moments of the GPDs and, with regard to their universality 
property, they provide the soft physics input for the calculation of 
hard wide-angle exclusive processes as for instance real Compton 
scattering, see Sect.\ \ref{sec8}.

Lattice QCD provides a method to calculate moments of the GPDs. In
fact the lowest three moments of the GPDs have been worked out as yet
\ci{haegler} in scenarios with pion masses between 350 and 800 MeV.
The extrapolation to the chiral limit cannot be performed with
a sufficient degree of accuracy as yet. In so far the comparison of the 
lattice results with experiment or other theoretical or
phenomenological results is to be done with reservation. 
Nevertheless, the $t$ dependencies of ratios of moments either
obtained from lattice QCD \ci{haegler} or from phenomenology
\ci{DFJK4} are surprisingly close each other for $-t\le 1.2\,\gev^2$
and for a large range of pion masses in the lattice calculation. 
Also the lattice results \ci{haegler} on the orbital angular momentum
are in fair agreement with \req{angular-momentum} given the
uncertainties in both the approaches. 

\begin{figure}
\begin{center}
\psfig{file=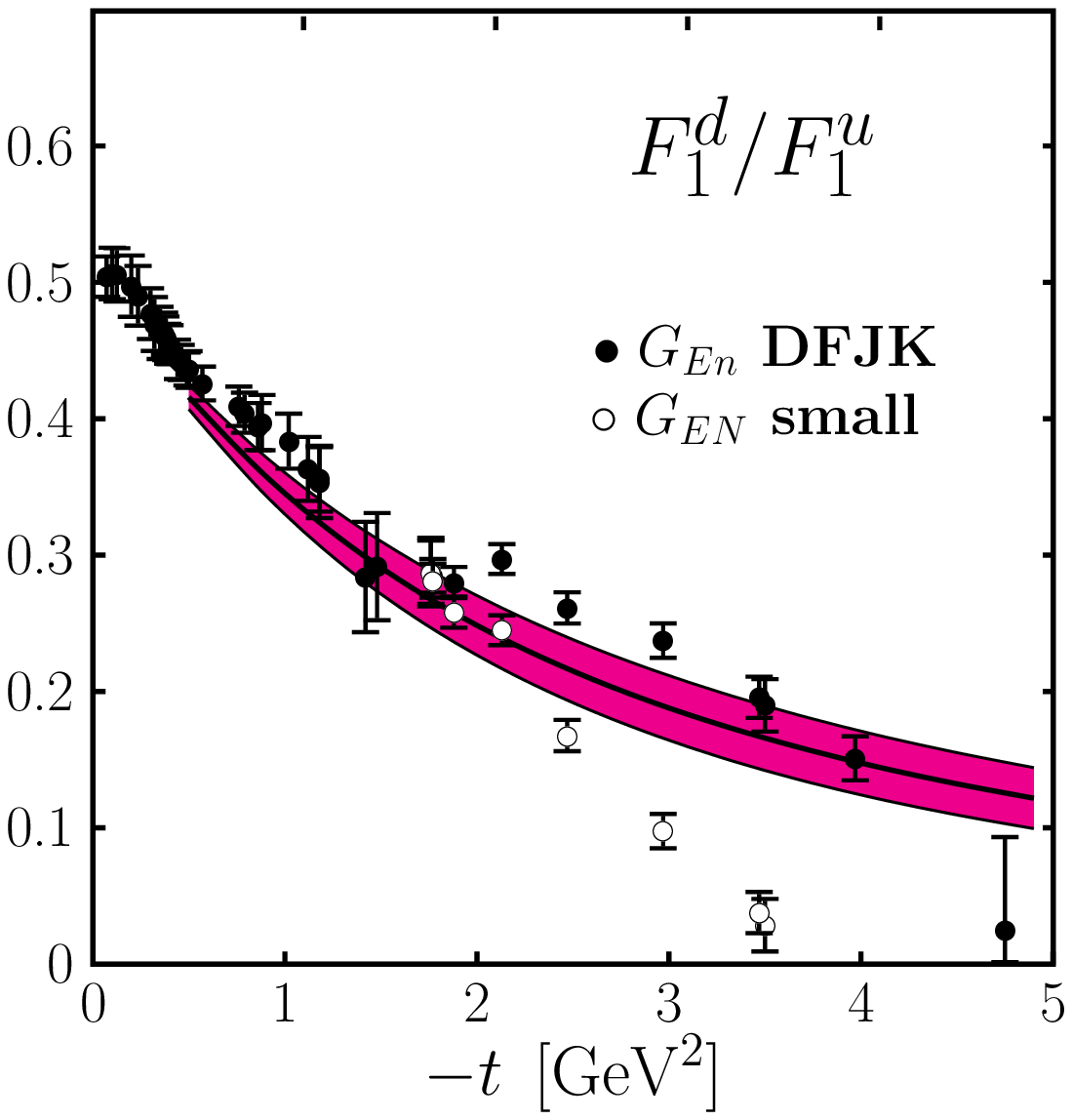,bb=105 363 425
  697,width=0.36\tw,clip=true}\hspace*{0.4\tw}

\vspace*{-0.42\tw}\hspace*{0.4\tw}
\psfig{file=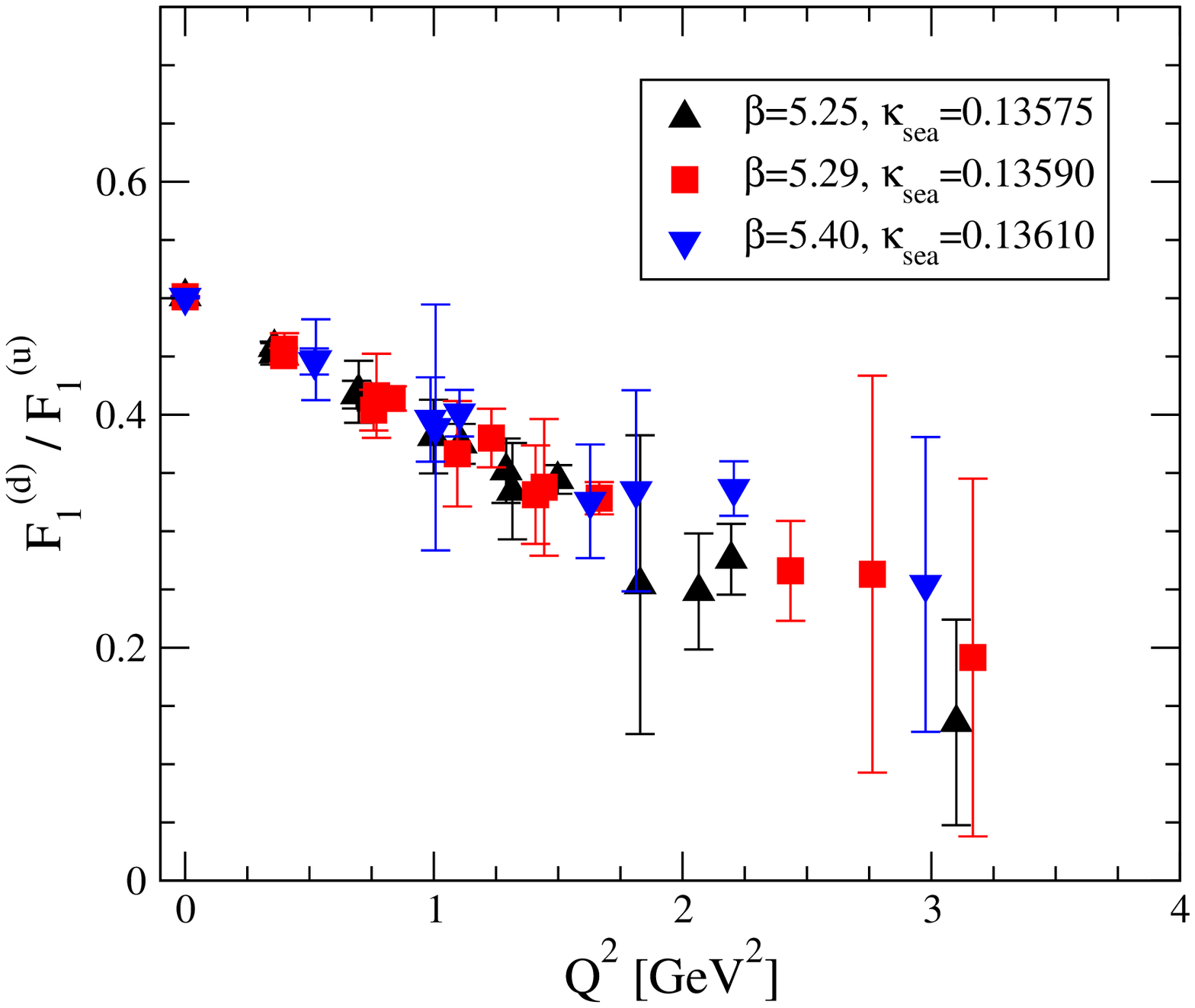,bb=147 22 576 528,width=0.37\tw,angle=-90,clip=true}
\end{center}
\caption{ The ratio of moments $F_1^d/F_1^u$ from Refs.\
  \refcite{DFJK4} (left) and \refcite{goeckeler06} (right) versus $-t=Q^2$.}
\label{fig:moments}
\end{figure}  
A particularly interesting feature of the GPD $H$  is that the ratio
of the lowest moments for $d$ and $u$ valence quarks, $F_1^d/F_1^u$,
drops rapidly with increasing $-t$, see Fig.\ \ref{fig:moments}. This 
feature is seen in both the phenomenological \ci{DFJK4} and the
lattice \ci{goeckeler06} analysis. It seems also to be demanded by 
experiment although presently an extrapolation of the neutron's
electric form factor is needed for the extraction of these moments
from data. The JLab Hall A collaboration (E02-013) will provide data
on $G_E^n$ up to about $3.5\,\gev^2$ in the near future which will
render an extrapolation unnecessary. Thus we have an indication that $u$
quarks may dominate over $d$ quarks in the proton form factor at large
$-t$, a behavior that corresponds to that of the PDFs at large $x$
\ci{cteq6}: $d_{\rm val}/u_{\rm val} \propto (1-x)^{1.6}$. This $t-x$
correlation of form factors and PDFs is a property of the ansatz 
\req{eq:ansatz}, \req{eq:profile}. Indeed one can show that the
moments of the form factors drop as
\be
         F_1^q \propto \mid t\mid^{-(1+\beta_q)/2}\,,
\ee
where $\beta_q$ is the power of $1-x$ with which the PDFs fall towards
$x=1$ (CTEQ6M \ci{cteq6}: $\beta_u=3.4$, $\beta_d=5$). These results
shed doubts on the assertion that the behavior of the Dirac form factor at
intermediate values of momentum transfer is a consequence of 
dimensional counting.   

\section{Deeply virtual exclusive scattering}\label{sec6}
\begin{figure}
\begin{center}
\psfig{file=handbag-c.ps,bb=118 236 342 408,width=0.25\tw,clip=true}
\psfig{file=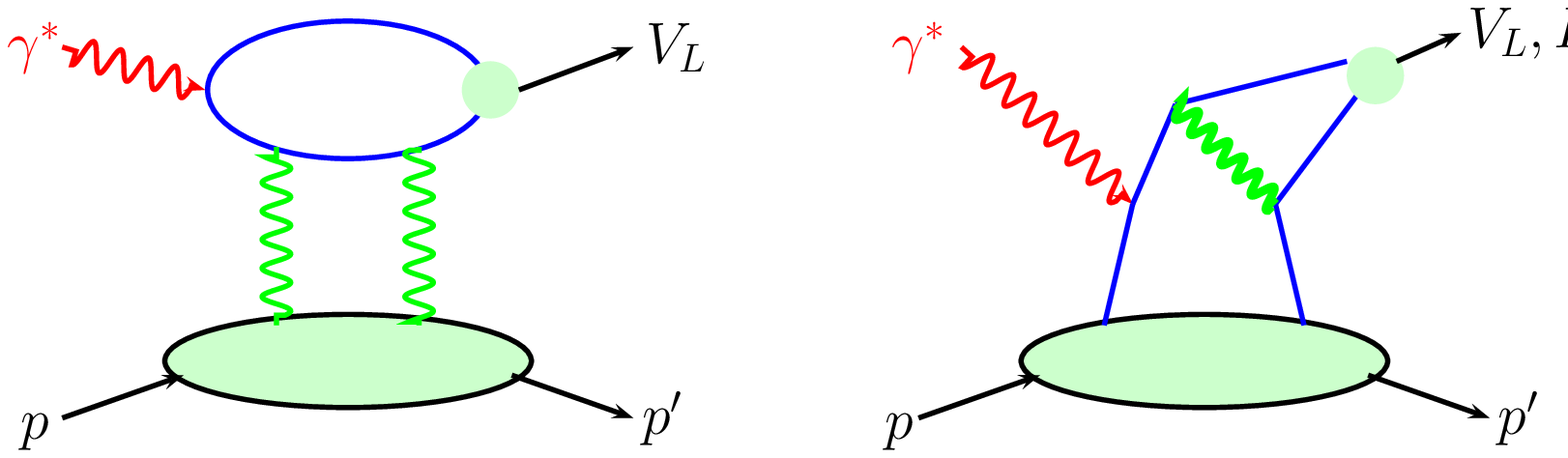,bb=125 495 605 655,width=0.56\tw,clip=true}
\end{center}
\caption{Typical graphs for deeply virtual electroproduction for
  $\gamma, V, P$.} 
\label{fig:graphs}
\end{figure}

Hard electroproduction of photons, vector mesons (V) and pseudoscalar
mesons constitute an important class of processes to which the
handbag factorization scheme can be applied to. In fact for these
processes rigorous proofs of factorization exist in the limit
$Q^2\to\infty$ \ci{rad96,ji97,collins}. In Fig.\ \ref{fig:graphs}
typical Feynman graphs are shown which contribute to these processes 
to leading-twist and LO pQCD accuracy. The dominant helicity
amplitudes ($\nu, \nu'$ ($\lambda, \lambda'$) label the helicities of 
the incoming and outgoing proton (parton), explicite helicities refer 
to those of photons and mesons) read
\ba
{\cal M}_{+\nu',+\nu}^\gamma &\sim& \sum_a
e_a^2 \int_{-1}^1 d\xb\left[ \frac{{\cal F}^a_{\nu'\nu} 
          +\widetilde{{\cal F}}^a_{\nu'\nu}}{\xb-\xi+i\epsilon}
+ \frac{{\cal F}^a_{\nu'\nu}-\widetilde{{\cal
	F}}^a_{\nu'\nu}}{\xb+\xi-i\epsilon} \right], \nn\\
{\cal M}_{0\nu',0\nu}^{M(q)} &\sim& \sum_a\,{\cal C}_V \int_{-1}^1\,
                   d\xb
  \left[ \sum_{\lambda} {\cal H}^{M(q)}_{0\lambda,0\lambda}\,
                          {\cal F}^a_{\nu'\nu} 
                 + \sum_{\lambda} 2\lambda {\cal H}^{M(q)}_{0\lambda,0\lambda}\, 
                 \widetilde{{\cal F}}^a_{\nu'\nu}\right],
\label{eq:DVES}
\ea 
where 
\ba
{\cal F}^a_{\nu\nu}=H^a-\frac{\xi^2}{1-\xi^2}\,E^a\,,  \qquad 
{\cal F}^a_{-\nu\nu}=2\nu \frac{\sqrt{t_0-t}}{2m(1-\xi^2)}\,E^a\,,
\ea
and analogously for $\widetilde{\cal F}$. For vector-meson production
there is an analogous contribution from the gluonic subprocess (see
Fig.\ \ref{fig:graphs}) to be added. Skewness is fixed in
electroproduction by Bjorken-$x$: $\xi\simeq \xbj/(2-\xbj)$ at small
$\xbj$. Since the interest lies in small $-t$, the $\gamma^*\to 
\gamma, V, P$ helicity non-flip transitions dominate. I.e.\ for the
Compton  process the transverse-transverse transition is leading while
the longitudinal-longitudinal transition obviously dominates for the
production of pseudoscalar mesons but also for vector mesons. This is
so since the subprocesses shown in Fig.\ \ref{fig:graphs}, suppress
transversely polarized vector mesons. Parity conservation tells us
furthermore that $\sum_\lambda \lambda {\cal H}_{0\lambda,0\lambda}=0$ 
for vector mesons while, for pseudoscalar mesons, $\sum_\lambda 
{\cal H}^{}_{0\lambda,0\lambda}=0$. In other words, electroproduction 
of pseudoscalar mesons probes the GPDs $\widetilde{H}$ and
$\widetilde{E}$, vector mesons the GPDs $H$ and $E$ to leading-twist 
order. To DVCS, on the other hand, all four GPDs contribute. 
The $t$ dependence of the subprocess amplitudes is usually neglected 
in contrast to that of the GPDs since it provides corrections of order
$t/Q^2$. With regard to flavor it is evident that DVCS probes the
valence and sea quark GPDs to LO pQCD, $\rho$ and $\omega$ production 
the gluon GPD in addition. The production of $\phi$ mesons is
sensitive to the gluon and sea GPDs, $J/\Psi$ production only to the 
gluon GPD since the charm content of the proton is tiny. The
production of $\pi^0$, on the other hand, is only fed by the valence
quark GPDs.

Different experiments probe different regions of $\xi$:
$\simeq 10^{-3}$ by HERA, $\simeq 10^{-2}$ COMPASS, $\simeq
10^{-1}$ HERMES and $\simeq 0.2-0.6$ JLab. Guided by the double
distribution model \req{DD}, \req{GPD-DD}, one expects that the role
of gluons and sea quarks is diminishing with increasing skewness
while that of the valence quarks is increasing. Thus, provided LO
handbag physics is dominant at a given hard scale, the study of the
mentioned processes over a wide range of $\xbj$ may allow to
disentangle the various GPDs.

Vector-meson electroproduction is dominated by the GPD $H$, the others
play a minor role. They are noticeable only in spin asymmetries like
$A_{LL}$ or $A_{UT}$ measured with longitudinally polarized
beam and target or a transversally polarized target, respectively.  
This is particularly the case for $\rho$ production, for $\omega$
production these effects are larger. Model estimates indicate
that for $\widetilde{H}$ and $E$ the valence quarks dominate for
$\xi\gsim 0.01$, sea quarks and gluons contributions seem to be small
and cancel each other to some extent. Now, for valence quarks the
following combinations occur
\be
F^\rho_v\=e_u F^u_v - e_d F^d_v \qquad
F^\omega_v\=e_u F^u_v + e_d F^d_v\,,
\ee
where $F_v=H_v, \widetilde{H}_v, E_v$. Given the signs of the GPDs
discussed in Sec.\ \ref{sec4}, we see that $H^\rho_v$ is large but   
$H^\omega_v$ is small while we have the opposite situation in the case
of $\widetilde{H}_v$ and $E_v$. Thus, $\omega$ production is probably
a very good case for studying $\widetilde{H}_v$ and $E_v$. This seems 
to be a rewarding task for JLab.  

\subsection{Deeply virtual Compton scattering}
\begin{figure}
\begin{center}
\psfig{file=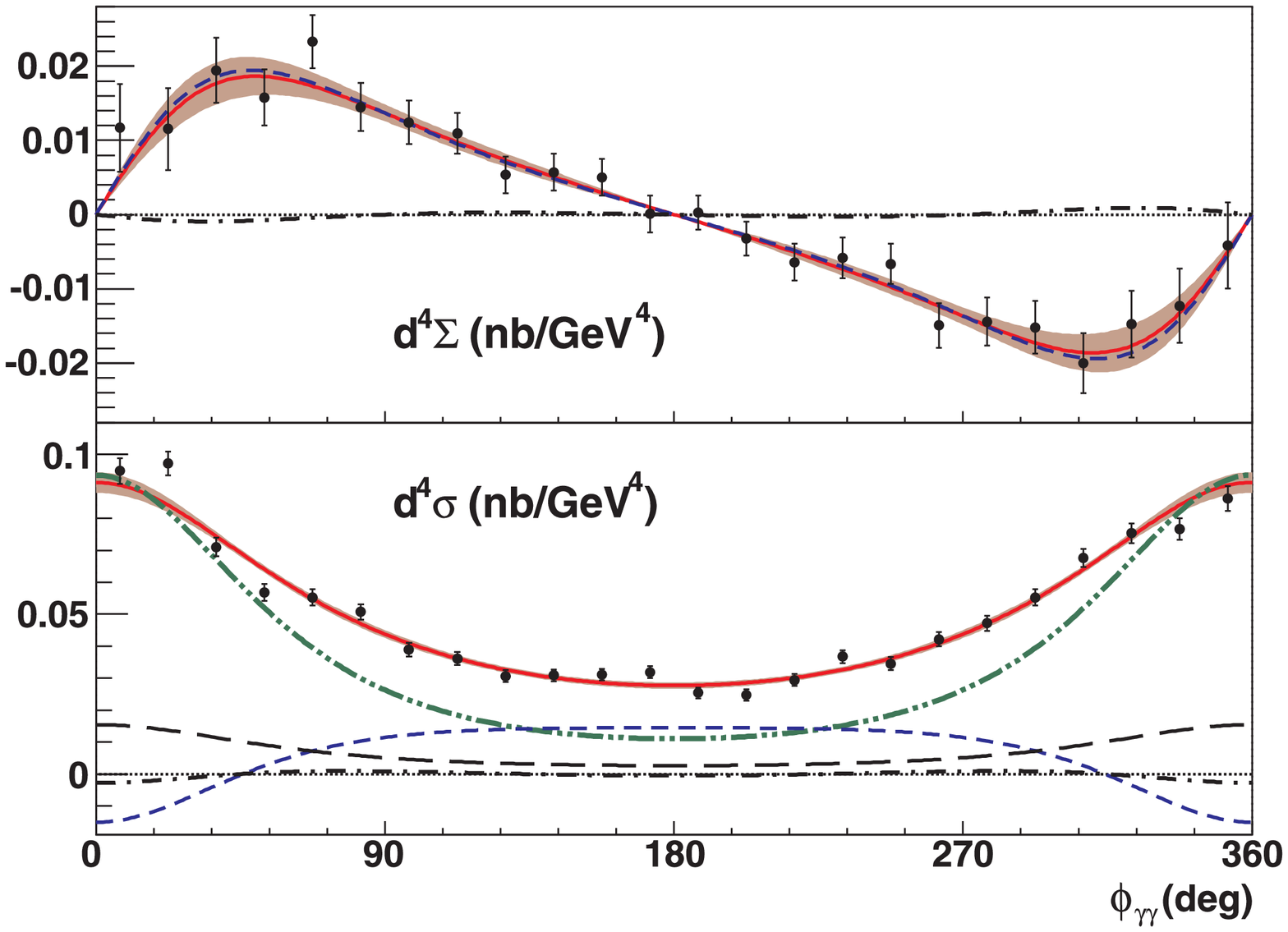,bb=18 138 583 379,width=0.74\tw,clip=true}
\end{center}
\caption{Left: The cross section for $ep\to e\gamma p$. The
  dash-dot-dotted line represents the Bethe-Heitler
  contribution. Data taken from Ref.\ \refcite{halla-dvcs}.}
\label{fig:dvcs}
\end{figure}

This process is considered to be the theoretical cleanest one and
therefore a lot of theoretical and experimental work is devoted to its
investigation. Still it is not a simple process. At NLO there are 
enhanced corrections from the gluonic GPDs which are particularly
large at low $\xi$ and overcompensate the suppression by $\als$ 
\ci{mueller06,ji98}. Another interesting feature of DVCS is the 
interference with the Bethe-Heitler process for which the final state
photon is emitted from the lepton. Since the Bethe-Heitler amplitude
is known for given nucleon form factors, the interference region of
both the contributions allows to study DVCS at amplitude level.
Measurements of $ep\to e\gamma p$ with a polarized beam or target
allows to filter out the interference term \ci{CLAS-P,halla-dvcs,HERMES-P}. 
In Fig.\ \ref{fig:dvcs} a recent result from the Jlab Hall A
collaboration \ci{halla-dvcs} is shown. Clearly seen are the 
interference regions and a region where the DVCS contribution
dominates. Whether the DVCS contribution seen in this experiment can
be understood within the handbag frame work is still a pending issue, 
detailed phenomenological analyses of DVCS data have not yet been 
performed. Only the HERA DVCS cross section data \ci{h1-dvcs,zeus-dvcs} 
have already been analyzed \ci{freund} to NLO with GPDs obtained from 
the double distribution model \req{DD}, \req{GPD-DD}. Recently methods 
have been developed that provide fitting schemes to the data by using
a kind of partial wave expansions of the DVCS amplitudes
\ci{mueller07}. These methods are not yet probed in detail. 

\subsection{Electroproduction of mesons}
\label{sec:dvem}
The disadvantage of meson electroproduction as compared to DVCS is
that a second soft hadronic matrix element is required, namely the
meson \wf{} or DA. This is to be traded for the advantage of separating
the GPDs $H$ and $E$ from $\widetilde{H}$ and $\widetilde{E}$ at
leading-twist accuracy. While there is a large set of accurate data 
available for vector meson electroproduction, only a few data exist as
yet for $\pi$ production \ci{hermes-pi}. Here I will restrict myself 
to a few comments on vector-meson electroproduction. 

\begin{figure}
\begin{center}
\psfig{file=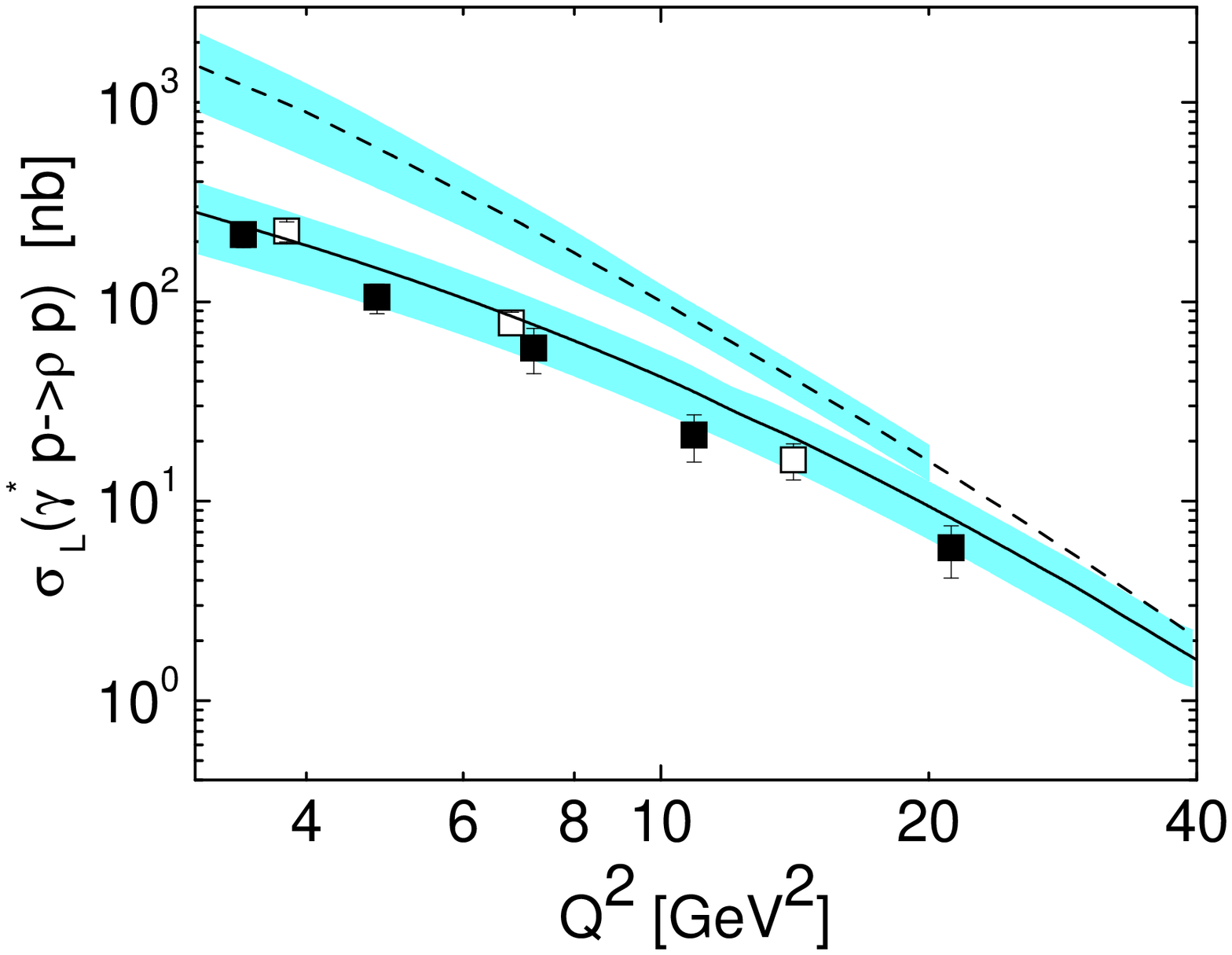,bb=37 313 534 700,width=0.45\tw,clip=true}\hspace*{1em}
\psfig{file=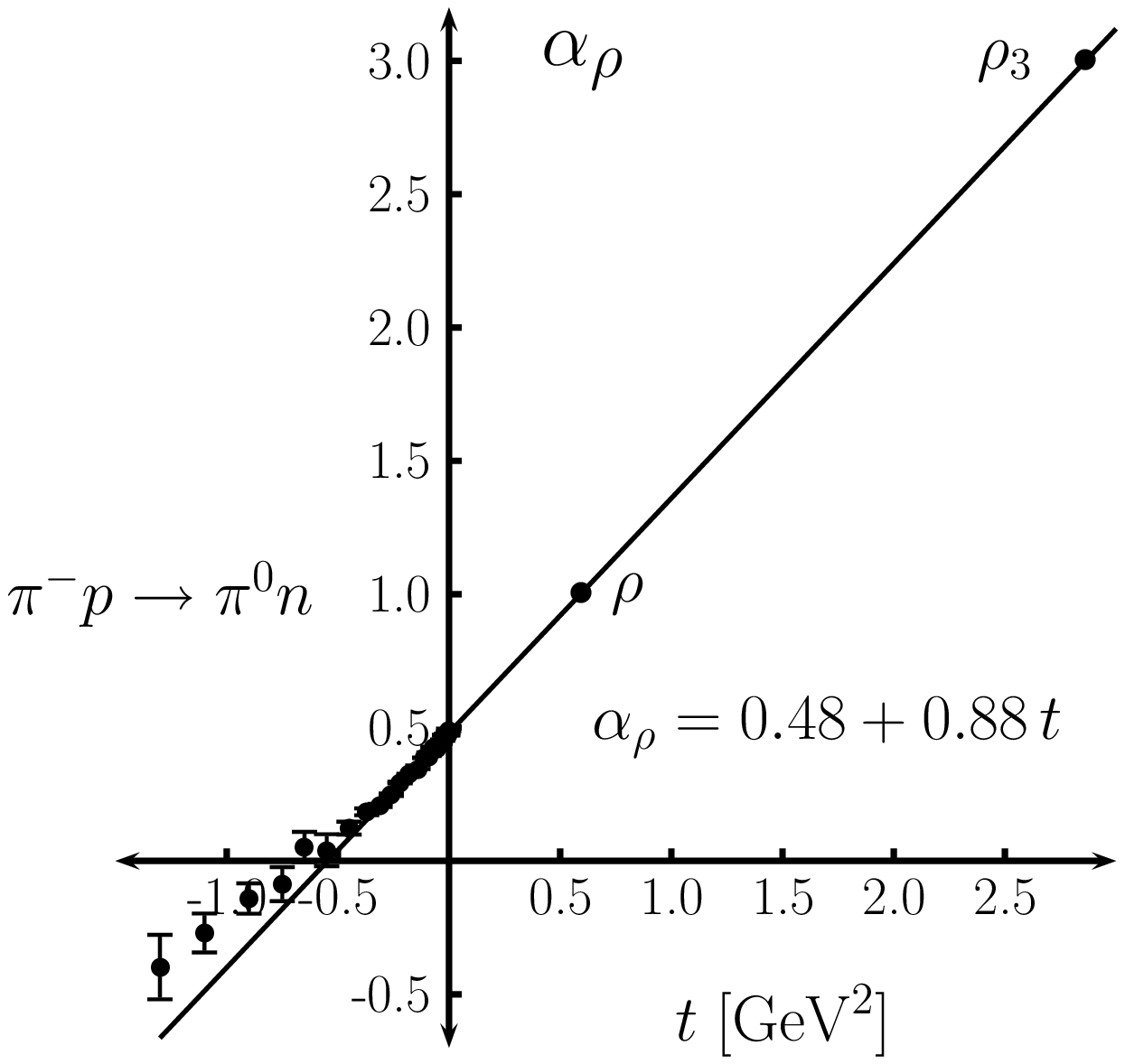,bb=100 313 465 656,width=0.38\tw,clip=true}
\end{center}
\caption{Left: The longitudinal cross section for $\rho$ production 
  versus $Q^2$ at $W=75\,\gev$. The solid line represents the handbag 
  predictions \ci{gk2}, the dashed line the leading-twist
  contribution. The bands indicate the theoretical uncertainties. Data 
  taken from Ref.\ \refcite{h1,zeus99}. Right: The $\rho$ Regge
  trajectory. Cross section data are taken from Ref.\ \refcite{charge-exch}.}
\label{fig:vector}
\end{figure}

There are several leading-twist, LO pQCD handbag calculations of 
vector-meson electroproduction \ci{VGG,diehl05,gk2} for which the
basic graphs are shown in Fig.\ \ref{fig:graphs}. It turned out
that the handbag contribution overestimates the cross section for 
longitudinal photons ($\gamma_L^*p\to V_Lp$) although with the tendency 
of approaching experiment with increasing $Q^2$, see Fig.\
\ref{fig:vector}. This an example of power corrections that persist 
up to very large scales.

It has recently been shown that NLO corrections \ci{ivanov04,kugler07}
are very large due to BFKL-type logarithms $\sim \ln{1/\xi}$ and
cancel to a large extent the LO term at low $Q^2$ and low $\xbj$. A
recent attempt \ci{ivanov07}  to resum higher orders with methods
known from DIS seems to indicate that the sum of all higher order 
corrections to the LO term is not large. Thus the issue of the size 
of higher order corrections is still unsettled.

A LO calculation that includes power corrections (modeled by quark
transverse momenta) in order to suppress the leading-twist
contribution to the $\gamma^*_Lp\to V_Lp$ amplitude and which also
allows to calculate the $\gamma^*_Tp\to V_Tp$ amplitude is advocated 
for in Refs.\ \refcite{gk2,golo3}. Only the subprocesses are
caluclated within the modified perturbative approach while the partons 
are still emitted and reabsorbed by the proton collinearly. The results 
for $\sigma_L(\rho)$ obtained in Ref.\ \refcite{gk2} are shown in
Fig.\ \ref{fig:vector}. With this approach good agreement with
experiment is also achieved for the ratio $R=\sigma_L/\sigma_T$, some spin 
density matrix elements, $A_{LL}$ and the target asymmetry $A_{UT}$.

Extension of this approach to other transitions is in principle
possible. Interestingly, while to the longitudinal amplitude only $H$ 
and $E$ contribute, the other amplitudes are also fed by $\widetilde{H}$ 
and $\widetilde{E}$. As shown in Ref.\ \refcite{golo1} the two types
of GPDs $H, E$ and $\widetilde{H}, \widetilde{E}$ lead to special
symmetry relations among the helicity amplitudes which are known from 
the exchange of particles with natural parity ($N$), and unnatural 
parity ($U$), respectively   
\be 
{\cal M}^{N(U)}_{-\mu'\nu',-\mu\nu}=
                   (-)(-1)^{\mu'-\mu}\, {\cal M}^{N(U)}_{\mu'\nu',\mu\nu}\,.
\ee
These symmetry relations prevent interferences between N and U type
contributions in unpolarized vector-meson electroproduction. Such 
terms however appear for instance in double spin asymmetries like $A_{LL}$. 

A final remark concerning vector-meson electroproduction is in order.
The cross section data \ci{clas04,hermes,h1,zeus99} reveal an asymmetric
minimum at $W \simeq 3 - 4\,\gev$ and fixed $Q^2$. The mild increase
of the cross section towards larger energies is well described by the
handbag physics but not the sharp increase in the opposite direction. 
Whether a new dynamical mechanism sets in at low $W$ or whether it is 
still handbag physics but with more complicated GPDs remains to be
seen. The upcoming data on $\rho$ electroproduction from CLAS may help 
in unravelling the physics in that kinematical region. 
 
\section{Alternative approaches to deeply virtual processes}
\label{sec7}
Vector-meson electroproduction has a long history. Its main
feature is that it behaves similar to elastic two-body reactions.
At the beginning this diffractive nature was understood with the
help of vector-meson dominance which views the photon as a
superposition of vector mesons and, hence, the process as elastic
vector-meson proton scattering. Pomeron exchange, supplemented by 
subleading Regge poles, lead to a fair description of vector-meson 
electroproduction at least at low photon virtualities. More
complicated versions of the Pomeron (soft and hard ones, BFKL Pomeron) 
allowed for an extension of the Regge model to larger values of $Q^2$. 
Later on the Pomeron was viewed as two gluons \ci{landshoff87} which 
couple perturbatively to the $q\bar{q}$ pair created by the virtual 
photon. Brodsky {\it et al} \ci{brodsky94} discussed the limit of 
large $Q^2$ but small $\xbj$ and showed that the Pomeron-proton vertex 
is approximately given by the gluon PDF $g(\xbj)$. This so-called 
leading-log$(1/\xbj)$ approximation which has frequently been applied 
\ci{frankfurt,martin,ivanov}. Similar to that approach is the 
color-dipole model \ci{nikolaev}. For the HERA setting of the 
kinematics, i.e.\ for $\xbj$ of the order of $10^{-3}$, the leading-log 
approximation is close to the handbag approach, the latter is only
enhanced by the skewness effect of about $20\%$. For larger values of 
$\xbj$ the leading-log approximation breaks down. It is also not clear 
how to generalize it to quarks. 

There is a renewed interest in Regge ideas, not only for vector-meson 
electroproduction and the small $x$ behaviour of the PDFs and GPDs but 
also for $\pi$ production and even for DVCS. Complete Regge fits to
data exist, e.g.\ Ref.\ \refcite{donnachie,laget}. The spectrum of hadrons
forms linear Regge trajectories (see Sect.\ \ref{sec4}) which means
that $J_j=\alpha_i(t=m^2_j)$ for a hadronic resonance with mass $m_j$
and spin $J_j$. The remarkable observation is that these Regge
trajectories, continued to negative $t$, describe the energy dependence
of the cross sections of soft two-body reactions at small $-t$. For 
instance, for $\pi^-p\to\pi^0n$ to which only the $\rho$ trajectory 
contributes, one finds
\be
d\sigma/dt(\pi^-p\to\pi^0n) \propto s^{2(\alpha_\rho(t)-1)}\,,
\ee
see Fig.\ \ref{fig:vector}. In other cases the cross section is
subject to a superposition of several Regge trajectories. 
To each Regge trajectory a residuum is associated which is a free
function of $t$. In spite of this interesting connection between the
particle spectrum and the energy dependence of cross sections, the
predictiveness of the Regge model is low. It often fails with
polarization observables but this can easily be cured by adding other
Regge poles and/or cuts. The Regge model lacks an important property any
theory and model should have - it cannot be disproved. 

\section{Wide-angle scattering} \label{sec8}
It has been argued \ci{DFJK1,rad98} that at large $s,-t,-u$ the
amplitude for real and virtual ($Q^2<-t$) Compton scattering
factorizes in analogy to DVCS (see Fig.\ \ref{fig:handbag-graph}). The
cross section for real Compton scattering reads in this case
\ba
\frac{d\sigma}{dt}&=& \frac{d\hat{\sigma}}{dt}\,\left\{
          \frac12 \Big[R_V^2(t)+\frac{-t}{4m^2}R_T^2(t)+R_A^2(t)\Big]\right.\nn\\
  &&\left.  -\frac{us}{s^2+u^2}
   \Big[R_V^2(t)+\frac{-t}{4m^2}R_T^2(t)-R_A^2(t)\Big]\right\}\,,
\label{eq:cross}
\ea
($d\hat{\sigma}/dt$ is the Klein-Nishina cross section).
Instead of a convolution as in \req{eq:DVES} $1/x$ moments of
zero-skewness GPDs occur now ($F_{vi}=H_v, \widetilde{H}_v, E_v$)
\be
R_i(t) \simeq \sum_{a=u,d} e_a^2\int_0^1\frac{dx}{x}\,F^a_{vi}(x,0,t)\,,
\ee
The tensor form factor $R_T$ describes proton helicity flip
\ci{huang}. With the zero-skewness GPDs \ci{DFJK4}, discussed in Sect.\
\ref{sec4}, at hand these Compton form factors can be evaluated and the
Compton cross section predicted; there is no free parameter. A very
good agreement with the recent JLab Hall A data \ci{halla-compton} is
achieved for sufficiently large Mandelstam variables. The handbag
approach also predicts interesting spin effects. For instance, the
helicity transfer from the initial photon to the outgoing proton reads
\be
K_{LL} \simeq \frac{s^2-u^2}{s^2+u^2}\,\frac{R_A(t)}{R_V(t)}\,.
\ee
Also this result is in agreement with a JLab measurement
\ci{hamilton}. The large positive value of $K_{LL}$ found in Ref.\
\refcite{hamilton}, is very difficult to achieve in the ERBL
factorization scheme \ci{ji06}.  

For the corresponding time-like process $\gamma\gamma\leftrightarrow
p\bar{p}$ a cross section, similar to \req{eq:cross}, can be derived 
\ci{DKV3,rad03} but the form factors are unknown and have to be
extracted from experiment \ci{belle}. It turns out that these form 
factors are larger than the corresponding space-like form factors, a
feature that is known from the electromagnetic form factors. The
handbag approach accounts for all the features of the BELLE data
\ci{belle}. It can be extended to other two-photon channels like
pairs of hyperons or mesons. It also applies to photoproduction of 
mesons and to $p\bar{p}\to \gamma M$. As for deeply virtual meson 
electroprodcution (see Sect.\ \ref{sec:dvem}) there are difficulties 
with the normalization of the cross sections \ci{huang-kroll} which 
have not yet been settled. Other features of these processes, as for 
instance the ratio of the $\gamma n\to \pi^-p$ and $\gamma p\to \pi^+n$ 
measured at Jlab \ci{zhu}, are quite well understood in the handbag 
approach.
 
\section{Summary}\label{sec9}
In this talk I have sketched the factorization schemes in use for hard
exclusive scattering processes and discussed their applications in
some detail. The main interest has been focussed on the handbag
approach since its prospects of becoming the standard description of
both the deeply virtual and the wide-angle exclusive processes are
best although a detailed comparison between theory and experiment is
still pending. With the exception of vector-meson electroproduction
for which already a vast amount of data exist, data for hard exclusive 
processes which cover a wide range of kinematics are still lacking but 
are excpected to become available in the near future from all
pertinent experiments. The upgraded JLab will provide even more data
on these processes in a few years. A definite judgement of the
handbag approach cannot be given at present. In case that the handbag
approach survives the detailed future tests we will learn much about
the GPDs and the structure of the proton. 

A special case are the valence quark GPDs at zero skewness which, with
a few assumptions, can be accessed through the data on the nucleon
form factors. JLab is in the position of providing more form factor 
data in the near future ($G_M^n$ from CLAS, $G_E^p$ from PR01-109, 
$G_E^n$ from E02-013) which will lead to improved GPDs. From the 
upgraded JLab more form factor data can be expected that will allow 
for an extension of the $t$ range in which the zero-skewness GPDs can 
be extracted. Lattice QCD results on moments of GPDs, provided these 
are reliably extrapolated to the limit of the physical mass of the 
pion, may diminish the dependence of these GPDs on the chosen 
parameterization.

\end{document}